# Observation of strong nonlinear interactions in parametric down-conversion of x-rays into ultraviolet radiation

**Authors:** S. Sofer[1*], O. Sefi[1*], E. Strizhevsky[1], S.P. Collins[2], B. Detlefs[3], Ch.J. Sahle[3], and S. Shwartz[1]

*S. Sofer and O. Sefi contributed equally to this work.

**Affiliations:**

[1] Physics Department and Institute of Nanotechnology, Bar-Ilan University, Ramat Gan, 52900 Israel

[2] Diamond Light Source, Harwell Science and Innovation Campus, Didcot OX11 0DE, United Kingdom

[3] ESRF – The European Synchrotron, CS 40220, 38043 Grenoble Cedex 9, France.

**Summary:** Nonlinear interactions between x-rays and long wavelengths can be used as a powerful atomic scale probe for light-matter interactions and for properties of valence electrons. This probe can provide novel microscopic information in solids that existing methods cannot reveal, hence to advance the understanding of many phenomena in condensed matter physics. However, thus far, reported x-ray nonlinear effects were very small and their observations required tremendous efforts. Here we report the observation of unexpected strong nonlinearities in parametric down-conversion (PDC) of x-rays to long wavelengths in gallium arsenide (GaAs) and in lithium niobate (LiNbO$_3$) crystals, with efficiencies that are about 4 orders of magnitude stronger than the efficiencies measured in any material studied before. These strong nonlinearities cannot be explained by any known theory and indicate on possibilities for the development of a new spectroscopy method that is orbital and band selective. In this work we demonstrate the ability to use PDC of x-rays to investigate the spectral response of materials in a very broad range of wavelengths from the infrared regime to the soft x-ray regime.

Optics and photonics is a major field of research that is important for fundamental sciences and has led to many practical applications and inventions. This field, which covers a broad range of the electromagnetic spectrum is, to a large extent, based on light-matter interactions. It is clear that a better understanding of the light-matter interactions is crucial for further developing of photonics-based research and technologies. Indeed, the approaches and techniques aiming at improving the knowledge of light-matter interactions are diverse and attract a great deal of attention[1]. However, most of the approaches utilize long wavelengths, thus cannot probe microscopic scale information, such as the valence electron redistribution in response to the application of external electric field or illumination. This is due to the fundamental limit of resolution of waves. Although x-ray-based techniques are capable of probing atomic scale structures, they interact mainly with core electrons thus provide very limited information on valence electrons and on their atomic scale interactions with long wavelength radiation[2]. There are several methods that utilize Bragg diffraction to reconstruct the valence charge density by subtracting the theoretical contribution of the core electrons from the measured intensity[3,4]. However, those methods are very limited, mainly since the largest contribution to Bragg diffraction is from the core electrons and therefore, the measurements require precision better than 1%[3]. Consequently, experimental results exhibit large variance and are often not reproducible[4]. Moreover, this method cannot give any spectroscopic information about the optical response of valence electrons.

About half a century ago, Freund proposed a method to measure the microscopic properties of valence electrons in solids by using nonlinear wave mixing of x-rays and long wavelengths such as UV and visible[5,6]. In essence, the effect can be viewed as x-ray scattering from optically modulated valence electrons, as is shown schematically in

Fig.1a, hence in this process the x-rays probe the variation in the state of the valence electrons and provide the access to the microscopic world. The advent of new x-ray sources such as the third-generation synchrotrons and more recently free-electron lasers has led to a substantial progress in the field. Previous works that studied x-ray and optical wave mixing exploited the effect of sum-frequency generation (SFG) and PDC[5-16]. The effect of difference frequency generation (DFG) of two x-ray pulses was studied theoretically as a probe for microscopic properties at the atomic scale[17,18]. Moreover, the effect of two photon absorption has been used recently for nonlinear spectroscopy[19]. The nonlinearity in x-ray and long wavelength mixing was discussed by numerous theoretical works[20-25]. However, to date, nonlinear interactions between x-rays and long wavelengths have been observed only in simple crystals such as diamond and silicon[8-15]. The typical efficiency and the signal-to-noise-ratio (SNR) that have been reported are very small and are barely sufficient for the measurements. The experimental results have been interpreted by using simple classical or semi classical models for the nonlinearity with reasonable agreements where the long wavelengths are either in the range where the crystals are transparent[11,14] or in the extreme ultraviolet regime with only one isolated resonance[8-10,12,13,15].

In this work we report on the observation of prominent large nonlinear effects in the non-centrosymmetric crystals GaAs and LiNbO$_3$, which are stronger by orders of magnitude from the background. Our measurements of the effect of PDC of x-rays into UV and the optical regimes in these crystals exhibit efficiencies that are about four orders of magnitude stronger than the efficiencies measured in any crystal, in which x-ray PDC has been observed so far. This is in sharp contrast to the prediction of all theories that have been used so far. Our results indicate an unrevealed underlying physical mechanism that is responsible for the strong nonlinearity we observed. In addition, we demonstrate the ability to perform spectroscopic measurements in a very broad range that enable the retrieval of information on the band structure, density of states, and atomic resonances in the crystals. As such, our work opens new and exciting possibilities for future research of nonlinear x-ray optics in complex materials and for the development of novel spectroscopy techniques that rely on these effects.

In the effect of PDC of x-rays into longer wavelengths, an input x-ray pump beam interacts with the vacuum fluctuations in the nonlinear crystal to generate photon pairs at lower frequencies. Energy conservation implies that the wavelength of one of the generated photons is in the x-ray regime (denoted as signal) and the wavelength of the second photon is in the UV or visible range (denoted as idler) where the sum of the frequencies of the generated photons is equal to the frequency of the input x-ray photon, namely $\omega_p = \omega_s + \omega_i$. Due to the high absorption in the UV range, the UV photons are completely absorbed and only the x-ray photon can be detected. However, since the photons are always generated in pairs, the rate of the x-rays depends also on the optical properties of the material in the UV range, hence the measurement of x-rays is sufficient to retrieve the information that can be probed by the UV photons. The selection of the wavelengths of the generated beams is done by using the requirement for momentum conservation (phase matching) that imposes a relation, which depends on the refractive indices of the material, between the propagation angles of the beams and their photon energies. Since the wavelengths of the x-ray photons are on the order of the distances between atomic planes, we utilize the reciprocal lattice vectors to achieve phase-matching, which is given by $\vec{k}_p + \vec{G} = \vec{k}_s + \vec{k}_i$ where $\vec{k}_p$, $\vec{k}_s$, and $\vec{k}_i$ are the k vectors of the pump, signal, and idler beams, respectively and $\vec{G}$ is the reciprocal lattice vector.

We conducted the experiments described in this article on beamline ID-20 of the European Synchrotron Radiation Facility and on beamline I16 of the Diamond Light Source [26]. The schematic of the experimental setup is shown in Fig.1 (b). The setup is very simple and relies on a standard diffractometer. We used a crystal analyzer to select the photon energy of the detected outgoing photons and an area detector to measure the profile of the scattered beam from the analyzer. The input beam is monochromatic and collimated. The samples we used are single crystals GaAs and $LiNbO_3$. We tuned the angles of the sample and the detector to the phase matching angles of the selected reciprocal lattice vector and photon energies. We rocked the sample and recorded the signal with the detector. Typical images of Bragg reflection and of the PDC signal are shown in Fig.1 (c) and (d) respectively. The Bragg reflection and the PDC signal are distinctive and are spatially separated from each other. From the images we construct the rocking curves of the sample (the signal count rate as a function of the sample angle) for various signal photon energies. At each of the energies we verify that we measure the PDC signal by comparing the positions of the peaks of the rocking curves to the calculated phase matching angles. We then reconstruct the spectra by registering the peak values of the rocking curves at each of the energies. Examples for the rocking curves can be found in the supplementary.

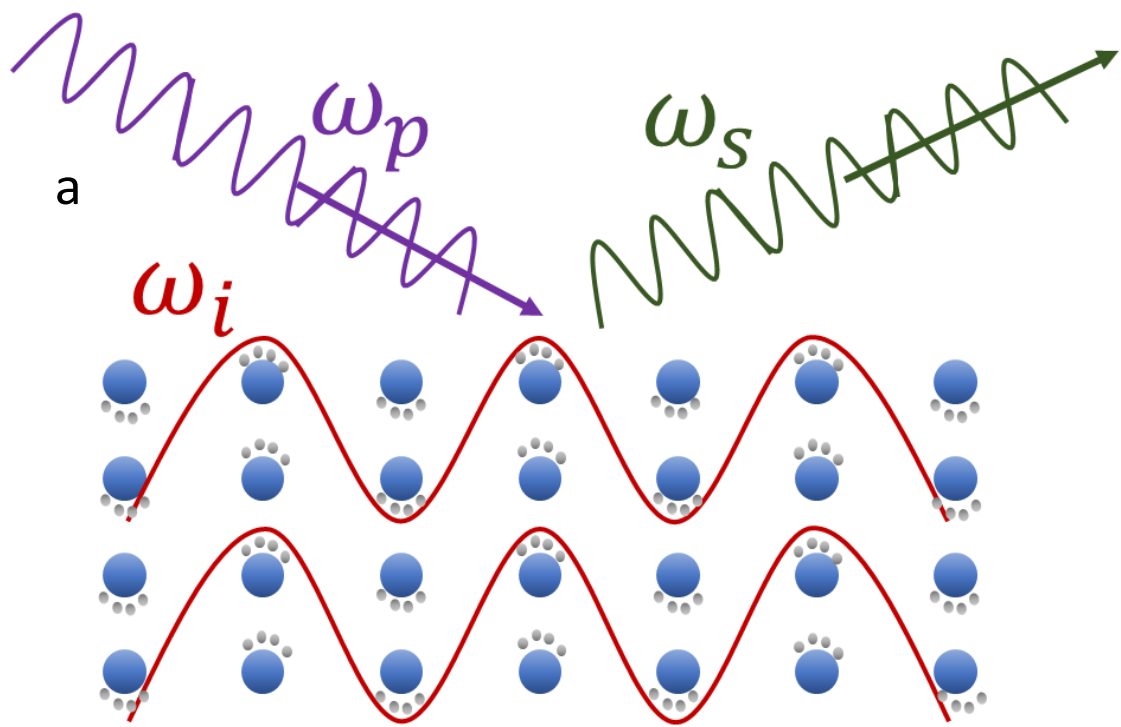

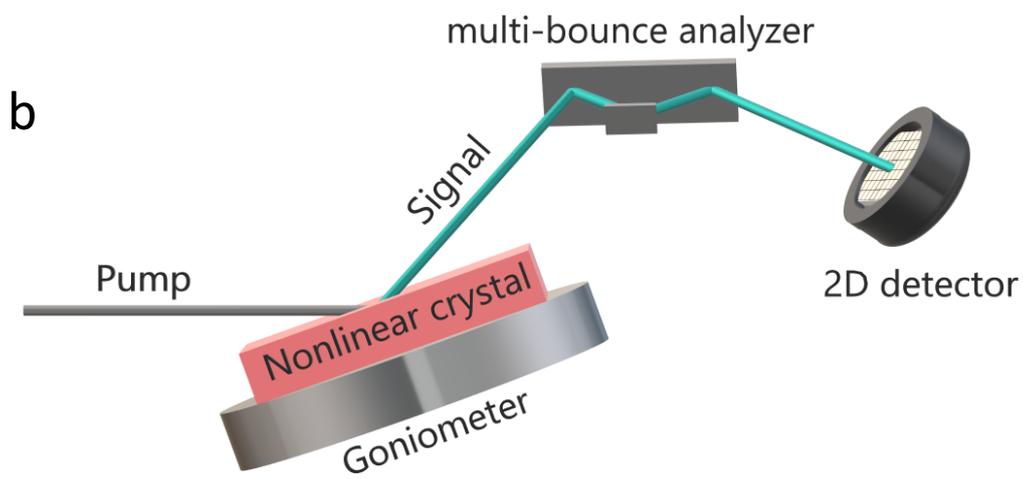

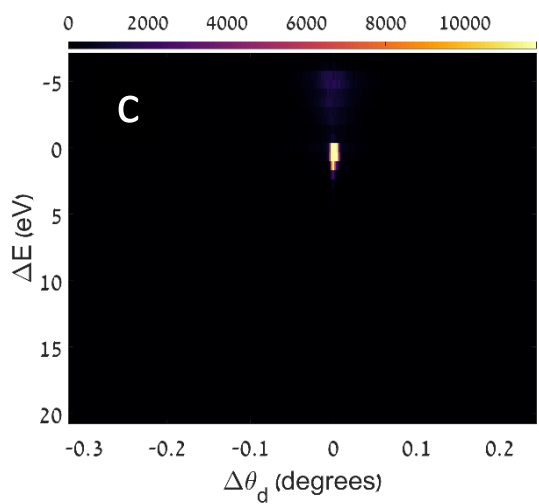

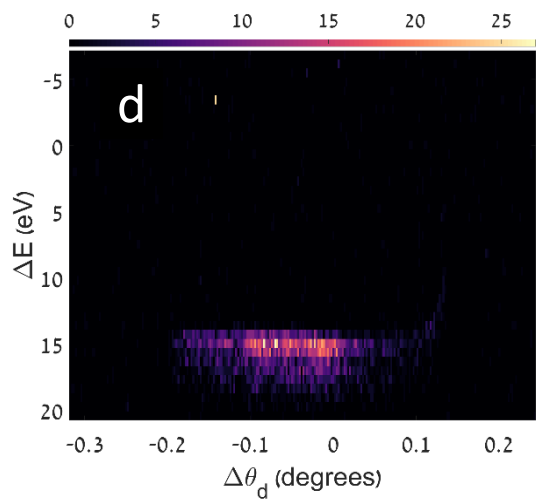

Fig 1: (a) Schematic description of the mechanism for the nonlinearity. The pump x-ray beam is scattered by an optically modulated charges and the frequency of the scattered x-rays is redshifted. (b) Scheme of the experimental setup. The synchrotron radiation illuminates a nonlinear crystal (GaAs or $LiNbO_3$). The PDC signal is selected by a multi-bounce analyzer and measured by a two-dimensional detector. The vertical axis of the detector is the deviation of the detected energy from the pump energy, which is determined by the analyzer angle. The horizontal axis is the angle with respect to the Bragg angle that corresponds to the pixel on the camera. (c) Typical image of Bragg reflection as taken by the detector. (d) Typical image of PDC. Note that while the profile of the Bragg reflection is very narrow in the horizontal direction the PDC profile is broad.

We begin by showing the very strong efficiencies of the signal of the PDC of x-rays into UV in both GaAs and LiNbO$_3$. Here we define the efficiency as the sum over the full width at half maximum of the PDC signal, as recorded by the detector, divided by the incident flux, while the sample is at the phase matching angle. The data analysis process is described schematically in the supplementary. We plot the efficiencies of the PDC for various idler energies and atomic planes in the GaAs and in the LiNbO$_3$ crystals in Fig. 2 and Fig. 3, respectively. In the figures, the abscissa is the inter-planar spacing and the vertical axis is the measured efficiency. The efficiencies range from $10^{-8}$ to $10^{-6}$, where in previous measurements of x-ray into UV PDC the efficiencies were on the order of $10^{-9} - 10^{-10}$ [13,14]. The general trend far from atomic resonances is that the efficiency drops as the idler photon energy increases. In addition, the theory, which led to a good agreement in the previous experiments [13-15], underestimates the efficiencies for GaAs and LiNbO$_3$ by about 4 orders of magnitude. This ratio is maintained in the whole range of measured photon energies. These pronounced discrepancies indicate on a new source for the strong nonlinearity, which was not considered in previous works.

Of importance, the measured efficiencies presented in Figs. 2 and 3 show a very different angular dependence from the angular dependence of the elastic scattering. Hence our results cannot be described by generalized scattering factors. The interpretation of the measured efficiencies for different directions of the crystal is essential for the understanding of the nonlinear interaction. The differences in the efficiencies for different atomic planes can be attributed to higher densities of valence electrons in different directions.

Another interesting result that is obtained from the measured efficiencies for LiNbO$_3$ is revealed when comparing reciprocal lattice vectors of nearly equal magnitude.

Although the x-ray scattering factors for these reciprocal lattice vectors are comparable, the efficiencies of the PDC effect for reciprocal lattice vectors, which are parallel to the c-axis of the crystal, are considerably larger. For example, the effect measured for the (0 0 6) atomic planes is two orders of magnitude larger compared to the (1 1 0) atomic planes. This can indicate that the nonlinearity depends on the direction of the permanent dipoles in the crystal and on symmetry of unit cells or the molecules in the materials. It is remarkable to note that for the visible range, the nonlinear coefficient is the largest for polarizations along the c-axis [27]. It is very likely that the electric field of the idler that leads to the largest efficiencies we observed in $LiNbO_3$ is also in the direction of the c-axis. This is consistent with the theories that predict that the largest efficiency for selected atomic planes is obtained when the electric field of the idler is parallel to the reciprocal lattice vector (normal to the atomic planes)[15].

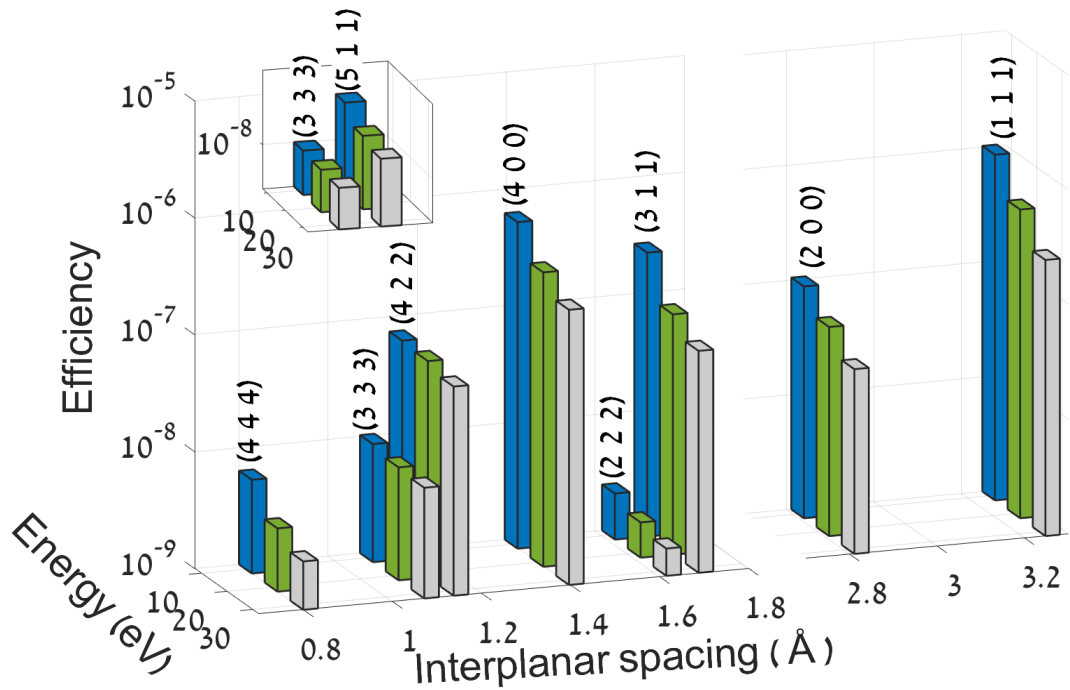

Fig.2: Comparison between measured efficiency of PDC in GaAs for several atomic planes for a bandwidth of 1 eV. The presented efficiencies are measured for signals, which correspond to idler energies of 10 eV, 20 eV, and 30 eV. The inset presents the efficiencies measured for two atomic planes with the same interplanar spacing. The horizontal axis represents the interplanar spacing, which corresponds to the Miller indices of the various atomic planes.

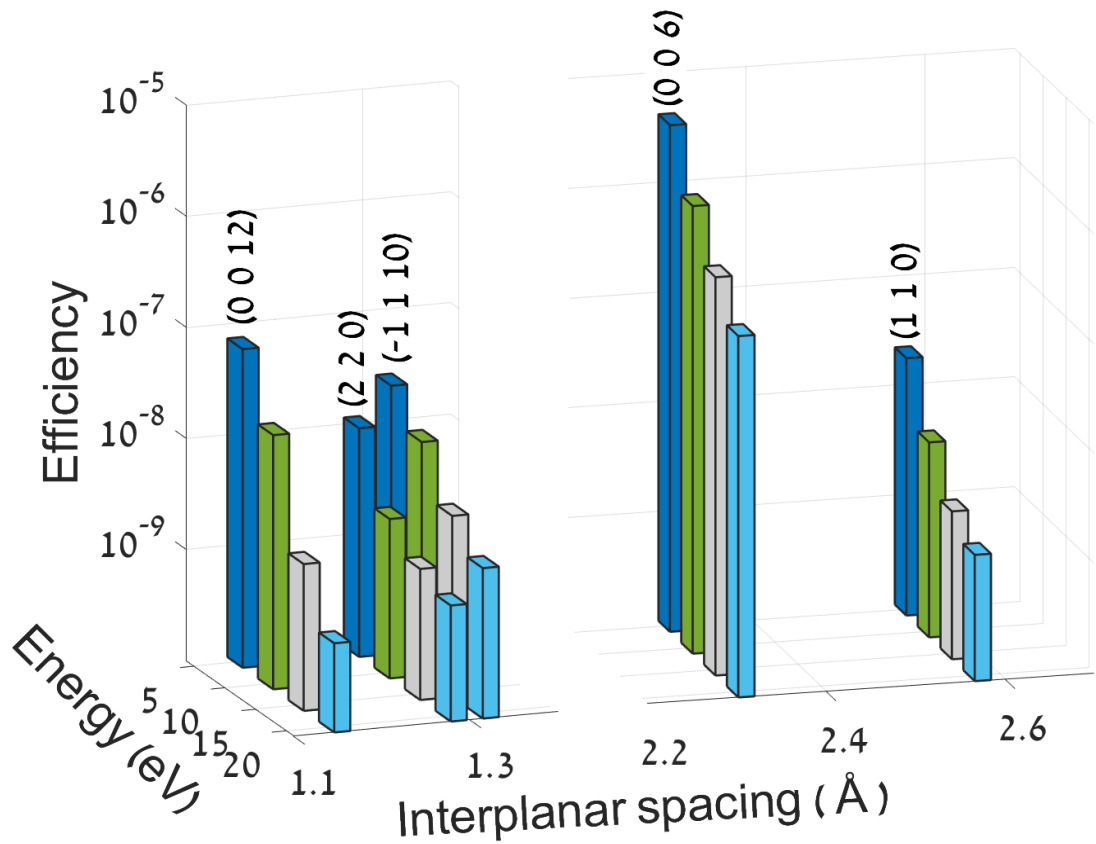

Fig.3: Comparison between measured efficiency of PDC in $LiNbO_3$ for several atomic planes for a bandwidth of 1 eV. The presented efficiencies are measured for signals, which correspond to idler energies of 5 eV, 10 eV, 15 eV, and 20 eV. The horizontal axis represents the interplanar spacing, which corresponds Miller indices of the various atomic planes.

After we show the observation of the very strong nonlinearities, we elucidate the origin of the effect and demonstrate the ability of its application for broad range spectroscopic measurements by exploring the spectral dependencies of the effect that are shown in Figs. 4 and 5 with a higher photon energy resolution. Fig. 4.a shows the spectrum measured for the GaAs (2 0 0) atomic planes. Here we show four clear features at idler energies of 8 eV, 17 eV, 28 eV, and 38 eV. We interpret the first two peaks as either direct transitions related to the band structure of GaAs, or atomic resonances of the arsenide atoms. The other two peaks cannot be attributed to band structure transitions (according to the present literature on the calculations of the band structure of GaAs[28]), but can be attributed to deeper atomic resonances of the arsenide atoms. The spectrum of the GaAs (4 0 0) atomic planes is shown in Fig.4 (b). Although this plane is parallel to the GaAs (2 0 0) planes, the measured spectra are clearly very different. This difference can be explained by considering the zinc-blende crystal structure of the GaAs, which suggests that the atoms in the GaAs (2 0 0) atomic planes are either gallium or arsenic atoms. This explanation is supported by observations in Fig. 4 (a) that show only atomic resonance of arsenide atoms. The peak that we observe for the GaAs (4 0 0) planes is interpreted as the transition of the direct band gap of GaAs. The measured spectra shown in Fig. 4 (c) and Fig. 4 (d) correspond to the GaAs (1 1 1) and GaAs (3 1 1) atomic planes, respectively. The spectra are still non-monotonic in the energy dependencies, with features that are above the noise level, but are less pronounced than the features shown in the previous spectra. The measured spectrum of the GaAs (1 1 1) atomic planes indicates that the highest efficiency is slightly above the band gap of GaAs as was measured for the (4 0 0) reflection, in addition to a feature at 6 eV, which can correspond to either transition at the band structure or to an ionization energy for the 4p valance electrons in gallium atoms. The spectra measured for the GaAs (3 1 1) again show a maximal efficiency slightly above the band gap. Moreover, we observe a sharp change in the trend of the spectrum near 20 eV. This feature is different than those discussed previously and cannot be interpreted as a resonance (from the band structure or atomic transitions).

In Fig. 5 the measured spectra of the $LiNbO_3$ crystal are shown. The peak at 5.9 eV for $LiNbO_3$ (1 1 0) and $LiNbO_3$ (3 3 0) atomic planes and the broad structures we observed for the $LiNbO_3$ (-1 1 10) atomic planes, which are shown in Fig.5 (a)-(c) respectively, can be attributed either to the direct band transition[29] or to the Li-2s atomic resonance. The peak at 7.5 eV, which appears in Fig. 3 (a)-(b) can be attributed either to the transition between the valence and the second conductance bands or to the Nb-4s atomic resonance. The peaks at 12 eV and 15 eV, which appears for the $LiNbO_3$ (-1 1 10) and the $LiNbO_3$ (3 3 0) atomic planes relate either to higher band transitions or to a deeper atomic level. Fig. 5 (d) shows the spectrum for the $LiNbO_3$ (0 0 6) atomic planes. In these measurements the atomic planes that participate in the effect are directed along the polarized direction on the crystal. This spectrum shows no prominent features.

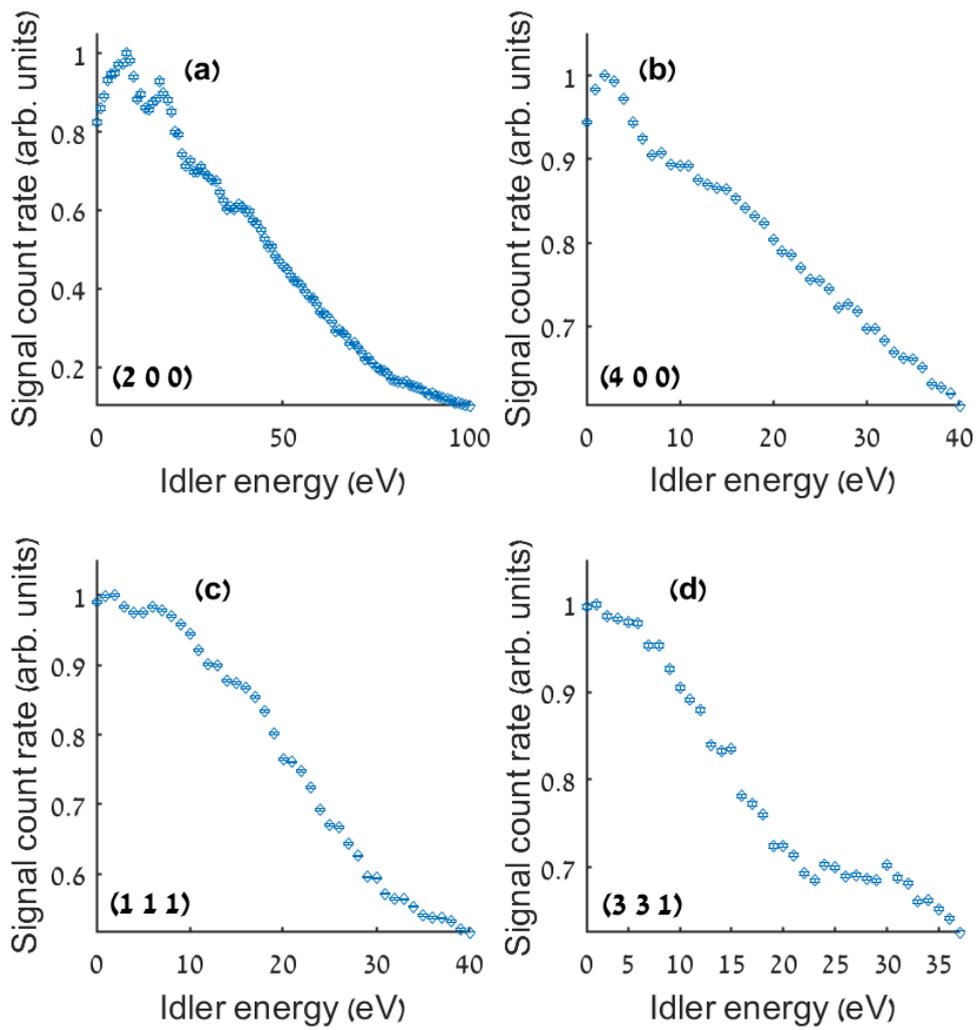

Fig 4: Spectral measurement of the PDC signal in GaAs for (a) the (2 0 0) atomic planes, (b) the (4 0 0) atomic planes, (c) (1 1 1) atomic planes, (d) the (3 1 1) atomic planes. The features in both subplots can be attributed to atomic resonances or to inter-band transitions in the crystal. See further details in the text.

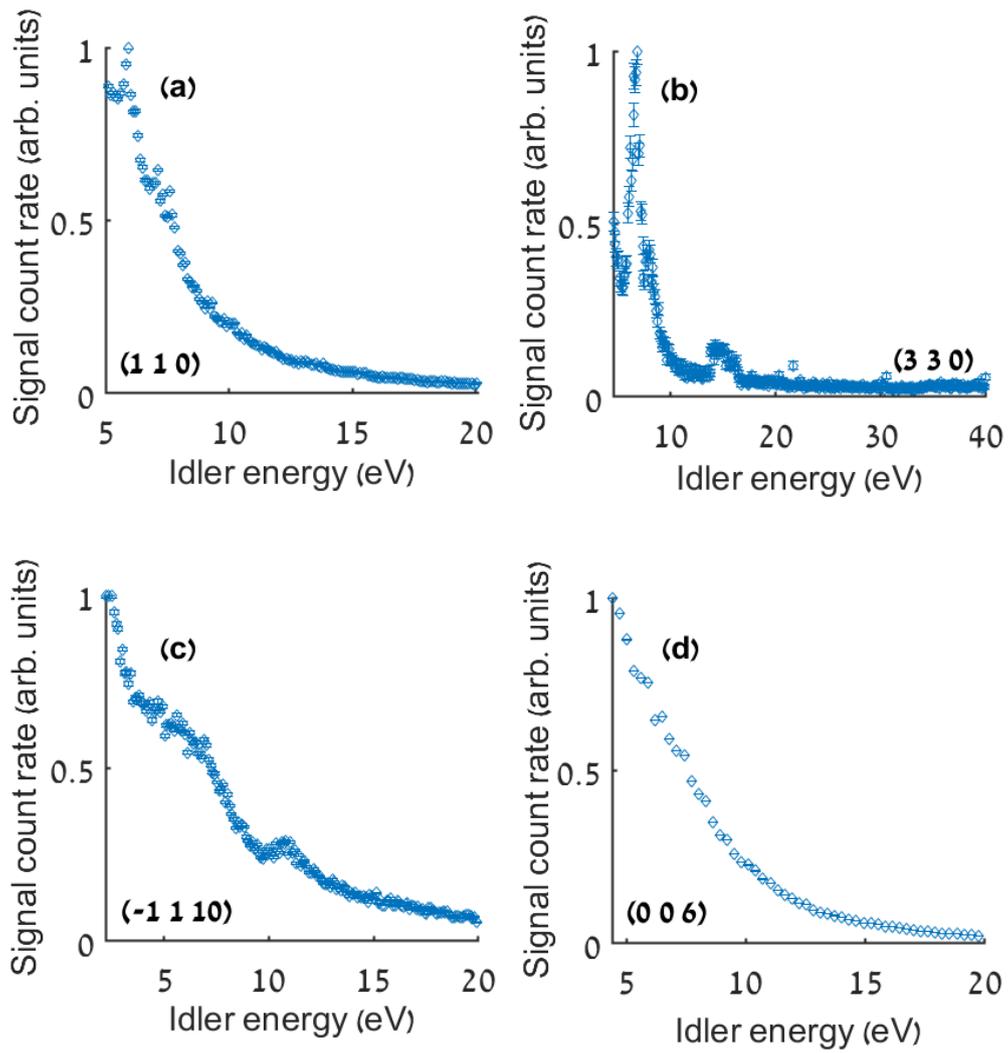

Fig.5: Spectral measurements for PDC signal in LiNbO$_3$ for the (a) (1 1 0) atomic planes, (b) (3 3 0) atomic planes, (c) (-1 1 10) atomic planes and (d) (0 0 6) atomic planes. The features in (a) – (c) can be attributed to either band transitions or atomic resonances. See further details in the text.

The main implication of the measured spectra is that the effect of PDC of x-ray into longer wavelengths can be used as a powerful tool to investigate phenomena in solid-state physics and in atomic physics. Generally speaking, it is likely that the lower idler energies correspond to solid-state physics and the higher energies to atomic physics, but to exactly distinguish these effects, a more detailed theoretical model is needed. Actually, the transition between the range where the dominant physical mechanism is solid-state physics and the range where the atomic physics dominates has been rarely studied. Hence, we foresee that PDC of x-rays into UV will be used as a powerful tool for studying this range, which will open an opportunity to study a broad range of physical phenomena with one experimental setup. We note that the peaks at the band gap of the crystal that appear in most of the measurements can be explained by the divergence of the joint density of states[25]. This suggests that PDC of x-rays into UV can be used for the study of electronic density of states and band structures in solids.

Another pronounced property that is common to all spectra is that apart from local enhancement near resonances, the efficiency decreases with the idler photon energies. However, the efficiency spectra are very different and depend on the atomic planes that are used for the phase matching. In addition, the efficiency dependencies on the idler energy in $LiNbO_3$ are stronger than in GaAs. This can be explained by noting that in $LiNbO_3$ there are fewer atomic resonances in the measured energy range. The different visible features in each of spectra can be explained by the nonlinearity, which depends also on the reciprocal lattice vectors themselves and on the distributions of the electronic wave functions. The comprehensive understanding of the spectra should include analysis of the nonlinear conductivity for different reciprocal lattice vectors while considering the selection rules between transitions.

We note that both inspected crystals have non-centrosymmetric crystal structures. This is in contrast to silicon and diamond crystals that have been investigated before. This observation raises the question of the importance of inversion symmetry to nonlinear effects with x-rays, which has been ignored in all previous publications. It is well known that in the optical regime the lack of inversion symmetry is essential for second order nonlinearities, however, thus far, x-ray nonlinearities have been considered as resulting from interactions beyond the diploe approximation that can be viewed as x-ray scattering from optically driven oscillations[11]. This type of second order nonlinearity is non-zero even in centrosymmetric martials but significantly weaker than the ordinary second order nonlinearity in non-centrosymmetric materials in the optical regime. It is therefore possible that the lack of inversion symmetry contributes significantly to the strong nonlinear effects we report here, which mixes between x-rays and long wavelengths. However, a comprehensive model that describes the interaction we measured in our experiments is still not available.

Finally, to further verify that our observations cannot be explained by inelastic scattering we scan the energy of the input beam in GaAs trough the absorption edges. We found that at the edges the PDC efficiency is reduced in contrast to inelastic scattering effects[30].

In conclusion, we have reported the measurement of non-predicted strong nonlinearities leading to efficiencies that are larger by about 4 orders of magnitude than previously reported efficiencies of PDC of x-rays into long wavelengths. These high efficiencies cannot be explained by existing theories and indicate an unexplored physical underlying mechanism. We note that the angular dependencies of the effect, the dependencies on the pump photon energy, and the agreement with the phase matching calculations constitute strong evidence that the effect we have measured is indeed PDC and not any other known inelastic effect. Our work demonstrates the possibility to utilize the effect of PDC of x-rays into visible and UV radiation as a powerful tool to probe numerous physical phenomena as it covers a large range of energies and provides structural and spectroscopic information in a single measurement. We expect that the full understanding of the effect will open a large range of opportunities to probe properties of solids and atoms, which are currently obscured due to the imitations of the present-day methods, hence advance the understanding of their functionality. Our work can be extended into the studying of the dynamics of the valence electrons and electronic excitations in a broad spectral range. This can be done by using short pulses generated by x-ray free electron lasers as the input beam for the PDC combining with an additional optical laser in a pump probe configuration.

**Acknowledgements:**

We acknowledge the European Synchrotron Radiation Facility for provision of synchrotron radiation facilities. We thank Diamond Light Source for access to Beamline I16. This work was supported by the Israel Science Foundation (ISF) (IL), Grant No. 201/17.

**Author contributions:**

O.S, S.Sofer, and S.Shwartz conceived the experiments. S.Sofer, O.S, E.S, B.D, S.P.C, Ch.J and S.Shwartz collected the data. S.Sofer and O.S preformed the data analysis. B.D, S.P.C, and Ch.J contributed to the experimental design and installation. S.Shwartz supervised the project. S.Sofer, O.S and S.Shwartz wrote the manuscript. All authors contributed to the work presented here and to the final paper.

## Methods

We provide further information on the experimental setups, which we used for both the GaAs and the LiNbO$_3$ experiments. First, we describe the experimental setup for the GaAs experiment, which we performed at Diamond Light Source on beamline I-16. The input power is approximately $10^{13}$ photons/sec. The input beam is monochromatic at 10.3 keV and is polarized in the scattering plane. The dimensions of the beam at the input are 20 μm (vertical) × 180 μm (horizontal). The nonlinear crystal is a GaAs crystal with thickness of 350 μm. The analyzer is a Si (1 1 1) triple-bounce analyzer and the detector is a Medipix multipixel detector with pixel size of 55 μm × 55 μm. The overall spectral resolution of the system is about 1 eV.

We performed the LiNbO$_3$ measurements at the ESRF on the ID20 beamline with a similar setup. The input power is approximately $5 \times 10^{12}$ photons/sec. The input beam is monochromatic at 10 keV and is polarized in the scattering plane. The dimensions of the beam at the input are 0.4 mm × 0.4 mm. The nonlinear crystal is a polished LiNbO$_3$ crystal with dimensions of 7 mm × 7 mm × 5 mm. The analyzer is a Si (4 4 0) double-bounce analyzer and the detector is a Medipix multipixel detector with a pixel size of 55 μm × 55 μm. The overall spectral resolution of the system is about 0.3 eV.

In both experiments the scattering plane of the analyzer was perpendicular to the scattering plane of the nonlinear crystal, in order to decouple the analyzer angle from the signal angle.

Next, we add further details on the data analysis procedure. We select the photon energy by tuning the analyzer angle. At each of the photon energies we scan the angle of the sample and record the intensity on the area detector for exposure times between 0.01 seconds and 1 second. At each angle of the sample we sum over the counts in the area within full width at half the maximum in the horizontal axis of the detector. We then plot the rocking curve (the dependence of the efficiency on the angle of the sample) and find its peak value. We repeat this procedure for the entire range of photon energies and plot the dependence of the peak value of the rocking curve as a function the photon energy to construct the spectra that we plot in Figs. 4 and 5.

To calibrate of the vertical pixels of the detector with respect to the idler energy, we use the following procedure. We tune the sample and the detector arm to the Bragg angles and the crystal analyzer to observe the highest intensity. This condition corresponds to the elastic scattering. We then find the vertical position on the detector where the Bragg signal is observed. By using the energy conservation equation $\omega_p=\omega_s+\omega_i$, we determine that this position corresponds to the idler energy of 0 eV, i.e. the elastic energy. Next, we detune the analyzer angle to fit for signal energy of 15 eV off the elastic energy. Energy conservation determines that this signal corresponds to idler energy of 15. We use linear interpolation to determine the idler energy along the vertical direction of the detector. The transformation of the horizontal position along the detector into degrees is performed as follows. We measure the Bragg horizontal position. This position is set to be the calculated Bragg angle. Next, we calculate the deviation from the Bragg angle by dividing the horizontal distance along the detector by the distance of the detector from the crystal. We repeat the calibration procedure for each of the atomic planes.

# Supplementary Material for "Observation of strong nonlinear interactions in parametric down-conversion of x-rays into ultraviolet radiation"

Authors: S. Sofer[1*], O. Sefi[1*], E. Strizhevsky[1], S.P. Collins[2], B. Detlefs[3], Ch.J. Sahle[3], and S. Shwartz[1]

*S. Sofer and O. Sefi contributed equally to this work.

**Affiliations:**

[1]Physics Department and Institute of Nanotechnology, Bar-Ilan University, Ramat Gan, 52900 Israel

[2] Diamond Light Source, Harwell Science and Innovation Campus, Didcot OX11 0DE, United Kingdom

[3] ESRF – The European Synchrotron, CS 40220, 38043 Grenoble Cedex 9, France.

We provide further description on our procedures for the data analysis and for the validation of the measurements of the PDC signal. We first describe our procedure for the reconstruction of the rocking curves and the spectra. As was mentioned in the main text, the spectra are reconstructed by finding the peaks of the rocking curves (the count rate of the signal as a function of the angle of the sample) at each of the photon energies of the idler (the longer wavelength photon). All rocking curves are generated by choosing a region of interest on the detector and summing over the counts within this region as a function of the detuning of angle of the sample from the Bragg angle. The regions of interests are chosen to be centered at about 20 pixels from the spot of the elastic scattering in the horizontal direction to block the strong signal from the residual elastic scattering. In the vertical direction they are centered around the peak. The sizes of the region of interest are about 50 pixels in the horizontal direction and 20 pixels in the vertical direction. The peak of each rocking curve is used to reconstruct the spectral dependence of efficiency of the PDC for the chosen atomic planes. A scheme for the reconstruction of the spectrum for chosen atomic planes is shown in figure S1.

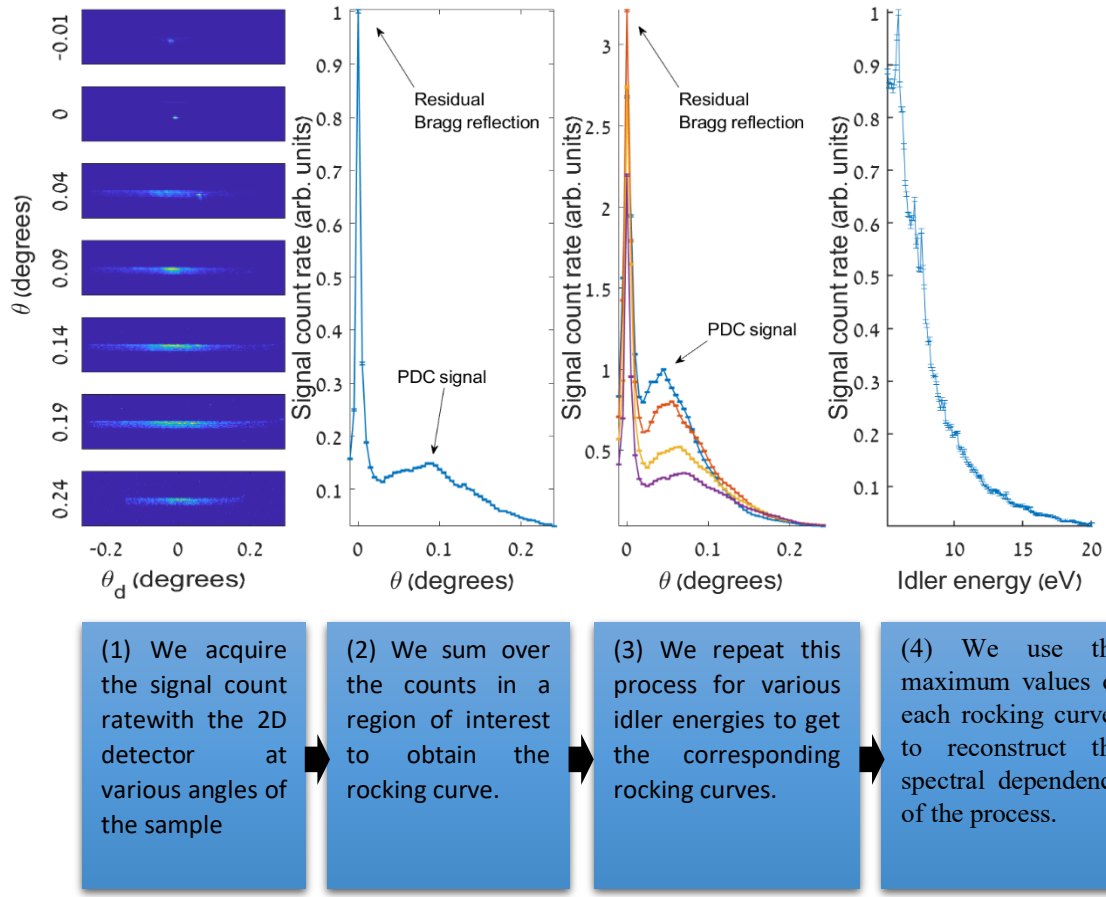

Fig. S1: Scheme of the data procedure. $\theta_d$ is the deviation of the signal from the Bragg angle and $\theta$ is the deviation of the crystal angle from the Bragg angle.

For the estimation of the efficiencies we take the peak of the rocking curve for the chosen atomic planes and sum over a region of interest that is defined by the full width at half the maximum of the signal PDC counts on the camera (while filtering any residual elastic scattering by removing them from the region of interest).

Next, we show several examples of such measured rocking curves that show the agreement with the calculated phase matching angles and nonlinearity, thus constitute conclusive evidence that the measured signal is indeed PDC. We recall that the angular dependence of the PDC efficiency is determined by the phase matching condition and by the nonlinearity. Thus, we expect that the maximal efficiency of the effect will be observed near the phase matching angles and when the nonlinearity is the largest. The equation that describes the phase matching is $\vec{k}_p+\vec{G}=\vec{k}_s+\vec{k}_i$. In order to estimate the shift from the phase matching condition, we use the phase mismatch, which is defined as $\Delta k_z L$ where $\Delta k_z$ is the deviation from the phase matching condition in the propagation direction and L is the length of the crystal, which is defined by the shortest absorption length. We use the wave vectors at the peak of the rocking curve to evaluate the phase mismatch. In the case of PDC of X-rays into UV radiation, the short length is the absorption length of the UV wavelength. The phase mismatch describes the dependence of the efficiency of the PDC process on the deviation from the exact phase matching condition.

We start by showing the rocking curves of $LiNbO_3$ for the (1 1 0) atomic planes at idler energy of 5 eV in Fig. S2. The angle of the pump wave vector at the peak of the rocking curve deviates by 0.06 degrees from the measured Bragg angle and the offset of the signal wave vector from the measured Bragg angle is 0.0225 degrees. The calculated mismatch is -0.575, which is much smaller than $\pi$.

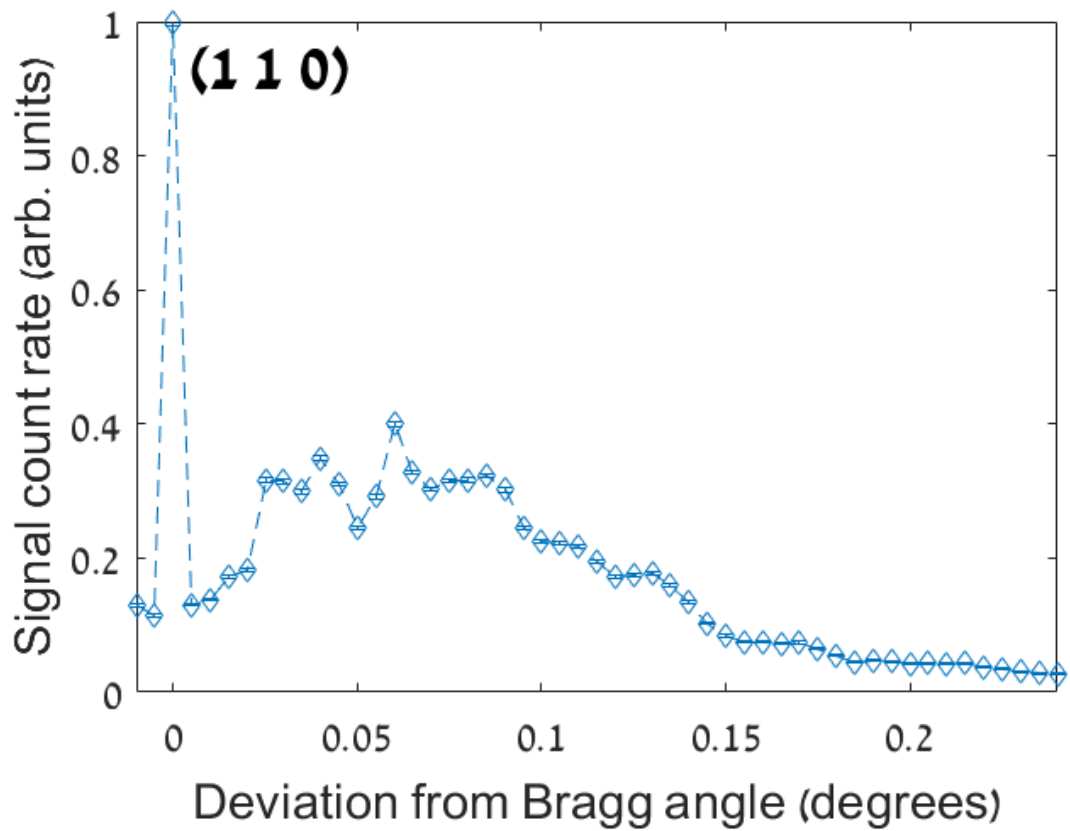

Fig. S2: LiNbO$_3$ rocking curve for the (1 1 0) atomic planes at idler energy of 5 eV. The PDC signal is the broad peak whereas the narrow peak on the left is the residual elastic scattering

We continue with the rocking curve for theLiNbO$_3$ (0 0 6) atomic planes for idler energy of 5 eV in Fig S3. The angle of the pump wave vector at the peak of the rocking curve deviates by 0.006 degrees from the measured Bragg angle and the offset of the signal wave vector from the measured Bragg angle is 0.0225 degrees. The calculated mismatch is -0.9789, which is smaller than π. The rocking curve is shown in Fig. S3.

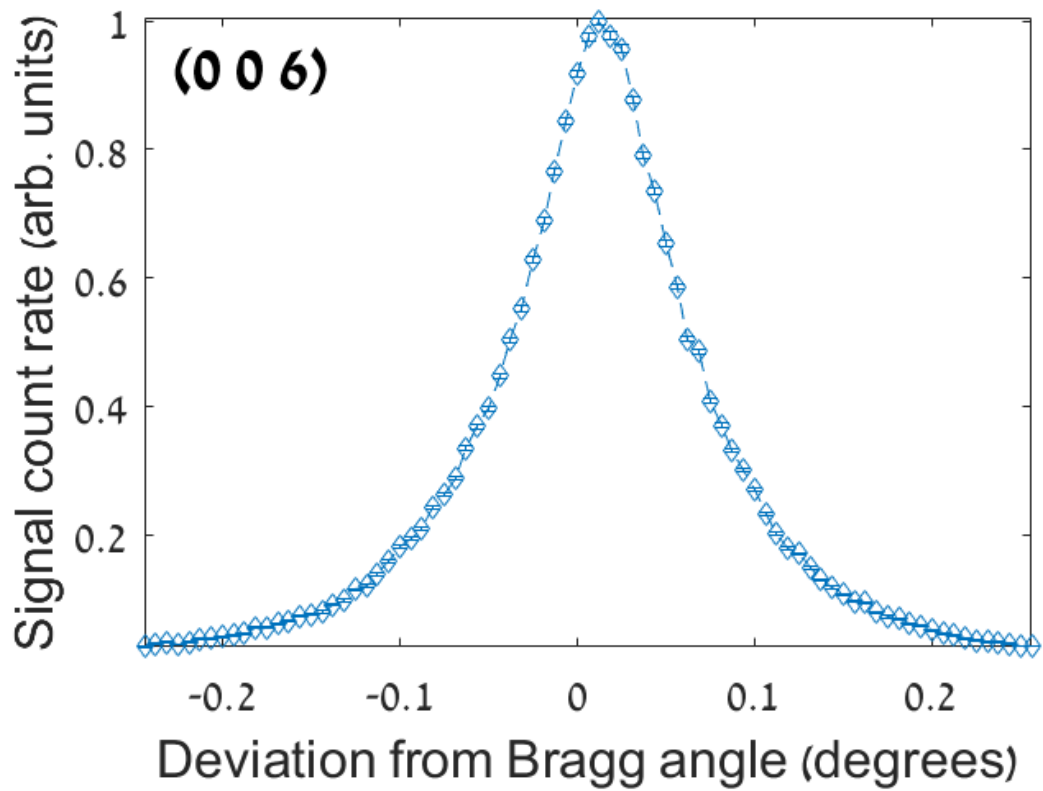

Fig. S3: LiNbO$_3$ rocking curve for the (0 0 6) atomic planes at idler energy of 5 eV.

Next, we show examples for rocking curves of GaAs. Fig. S4 shows the rocking curve measured for the (4 0 0) atomic planes for idler energy at 20 eV. The angle of the pump wave vector at the peak of the rocking curve deviates by 0.08 degrees from the measured Bragg angle and the offset of the signal wave vector from the Bragg angle is -0.01 degrees. The calculated phase mismatch is -1.221, which is also smaller than $\pi$.

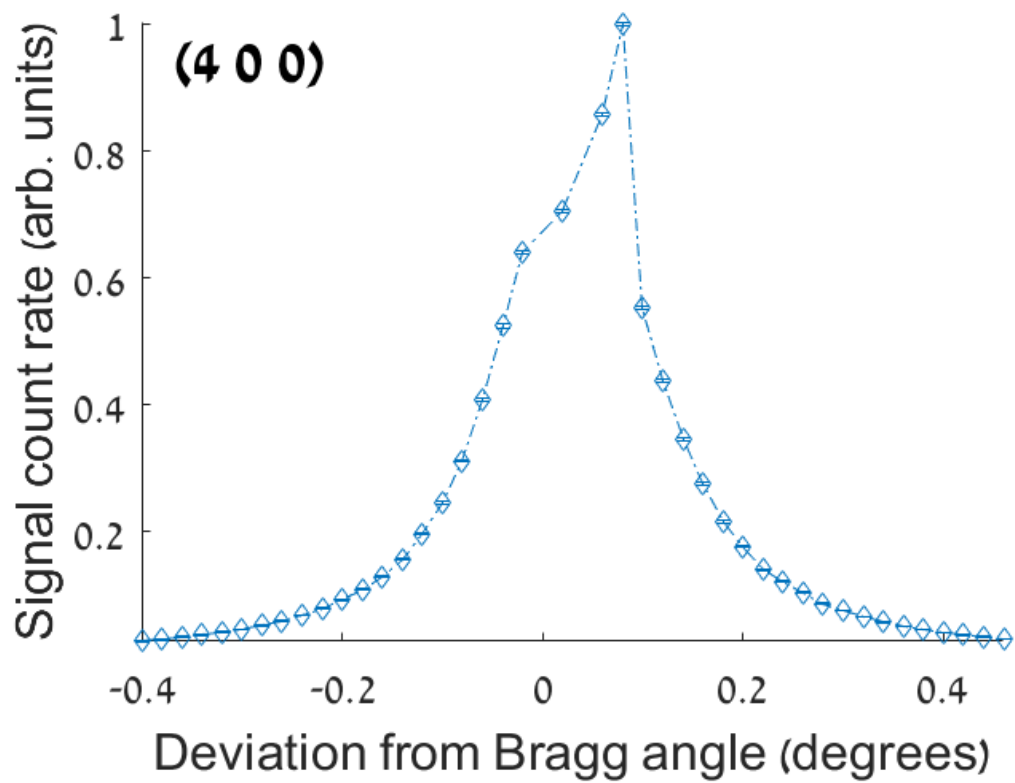

Fig S4: GaAs rocking curve for the (4 0 0) atomic planes at idler energy of 20 eV.

We next show the rocking curve measured for the (3 1 1) atomic planes in Fig. S5. The angle of the pump wave vector at the peak of the rocking curve deviates by 0.14 degrees from the Bragg angle and the offset of the signal wave vector from the Bragg angle is -0.023 degrees. The calculated phase mismatch is -0.1, which is again much smaller than $\pi$.

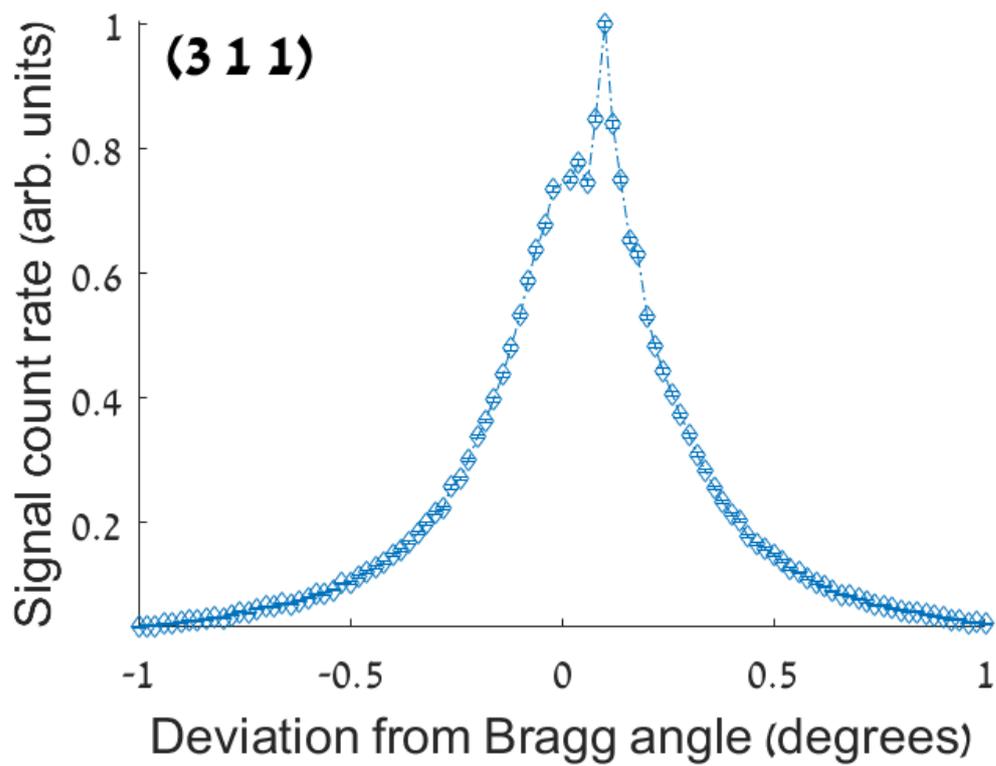

Fig S5: GaAs rocking curve for the (3 1 1) atomic planes at idler energy of 20 eV

We note that in addition to the uncertainties regarding the short UV absorption lengths, there are also experimental uncertainties that emerge from the bandwidth and the acceptance angles of the input monochromator and the analyzers. We estimate the overall energy precision to be 0.3 eV at ESRF and 1 eV at the Diamond Light Source.